\documentclass[aps,superscriptaddress,twocolumn,nofootinbib]{revtex4}



\usepackage{natbib}

\usepackage{graphicx}
\usepackage{dcolumn}
\begin{document}


\title{Limit order placement as an utility maximization problem and the origin of power law distribution of limit order prices}

\author{Fabrizio Lillo}
\affiliation{Dipartimento di Fisica e Tecnologie Relative, 
Universit\`a di Palermo, Viale delle Scienze, I-90128 Palermo, Italy.}
\affiliation{Santa Fe Institute, 1399 Hyde Park Road, Santa Fe, NM 87501}

\date{\today}

\date{Received: date / Revised version: date}
%
\begin{abstract}
I consider the problem of  the optimal  limit order price of a financial asset in the framework of the maximization of the utility function of the investor. The analytical solution of the problem gives insight on the origin of the recently empirically observed power law distribution of limit order prices. In the framework of the model, the most likely proximate cause of this power law is a power law heterogeneity of traders' investment time horizons .
\end{abstract}  
\maketitle

\section{Introduction}

In the last years physicists have shown a considerable interest in the empirical investigation of statistical regularities in socioeconomic systems. The next step of this type of investigation should be  the understanding of the origin of the discovered statistical regularities. Ultimately the explanation should be given by modeling the behavior and the preferences of the agents playing in the system. However there is often a significant mismatch between the way of modeling how agents' preferences are matched and the real mechanism through which agents take their decision and meet other agents' preferences.
Consider for example agents in a financial market. The theory of supply and demand describes how prices fluctuates as the result of a balance between product availability at each price and the desires of buying at each price. However in any real financial market, agents refrain from revealing their full supply or demand, trade incrementally large orders, choose strategically the timing and the amount of supply or demand they put in the market. In other words the agent's decision making process takes into account the specific way in which market works.  Financial markets are an optimal system to study agents' preferences and decision making strategies and their effect on statistical regularities of financial time series. This is due to the large availability of detailed data and to the relatively simple market structure. Several papers have been published on the optimal strategy of traders but these studies consider mainly a specific type of trader, the dealer or liquidity provider \cite{Cohen,Ho,Bouchaud,Avellaneda}. A paper more related to the present one but that consider a simplified price dynamics is Ref.~\cite{Wald05}.
The market structure of most financial markets is the limit order book (or continuous double auction). The limit order book is a queue where the list of buy and sell limit orders are stored. A limit order is an order to buy or to sell a given amount of shares at a specified price or better. If there is no one in the market willing to sell or buy at these conditions, the limit order is stored in the book and the agent can wait until the price hits the limit price and the transaction occurs. Of course the trader has the freedom to cancel her limit orders when she wants to. The decision on the limit price and volume of the order is a typical case of decision making under risk. 
In this paper I consider the problem of limit order placement in the framework of decision making.

One of the statistical regularities recently observed in the microstructure of financial markets is the power law distribution of limit order price in continuous double auction financial markets \cite{Zovko,Potters}. Let $b(t)-\Delta$ denote  the price of a new buy limit order, and $a(t)+\Delta$ the price of a new sell limit order. Here $a(t)$ is the best ask price and $b(t)$ is the best sell price. The $\Delta$ is measured at the time when the limit order is placed. Different  authors have investigated the probability distribution for the quantity $\Delta$ in different financial markets. It is found that $p(\Delta)$ is very similar for buy and sell orders. Moreover for large values of $\Delta$ the probability density function is well fitted by a single power-law
\begin{equation}
p(\Delta)\sim\frac{1}{\Delta^{\zeta_\Delta}}
\label{pow}
\end{equation}
There is no consensus on the value of the exponent $\zeta_\Delta$. Farmer and Zovko \cite{Zovko} estimated the value $\zeta_\Delta=2.5$ for stocks traded at the London Stock Exchange, whereas Potters and Bouchaud \cite{Potters} estimated the value $\zeta_\Delta=1.6$ for stocks traded at the Paris Bourse. More recently Mike and Farmer \cite{Mike} fitted the limit order distribution for LSE stocks with a Student's distribution with $1.3$ degrees corresponding to a value $\zeta_\Delta=2.3$.
In their study Farmer and Zovko studied also the correlation between limit order placement and volatility. They found a significant simultaneous cross correlation between volatility and limit order placement, indicating that when volatility is high, traders tend to place limit orders with larger values of $\Delta$. Moreover a lagged cross correlation analysis indicates that volatility leads relative limit price.

In this paper I investigate the origin of this power law distribution. To achieve this goal it is important to model the way in which traders placing limit orders decide the limit price of their orders. Suppose that a trader wants to place a limit order to sell. If she choose a very high limit order price she potentially makes a large profit, but it is unlikely that the order is matched in a reasonable time. On the other hand if the limit order price is close to the actual best available sell limit price, the limit order is likely to be matched in a short time but the profit is small. The right limit order price is the result of a tradeoff between these two choices and it depends on the characteristics of the agent as well as of the market state. 

The paper is organized as follows. In Section II I introduce the problem of limit order placement and the modeling  of the decision making process. I also find the solution to the problem in some specific but important cases. In Section III I use the result of the optimization to investigate the origin of power law tail in the limit order price distribution. In Section IV I perform an empirical investigation to test the results of the model and Section V concludes.

\section{The utility maximization problem}

Expected utility is probably the most important theory of decision making under risk  \cite{Ingersoll}. An individual is faced with a prospect (or lottery) $(x_1,P_1;...;x_n,P_n)$ which is a contract that yields outcome $x_i$ with probability $P_i$, where $\sum_{i=1}^n P_i=1$. Expected utility theory states the existence of an utility function $u(x)$ and the expected utility of the prospect is 
\begin{equation}
U(x_1,P_1;...;x_n,P_n)=\sum_{i=1}^n P_i u(x_i).
\end{equation}
When confronted with more prospects, the individual chooses the one which maximizes the expected utility $U$. Different individuals may have different utility functions $u(x)$. Usually the function $u(x)$ is concave, i.e. $u''<0$ which implies risk aversion. A very common utility function is the constant relative risk aversion utility function
\begin{equation}
u(x)=C~x^{\alpha}
\label{exp}
\end{equation}
In this equation $\alpha$ measures the level of risk aversion, which is larger for smaller $\alpha$. The value $\alpha=1$ describes a risk neutral individual, i.e. an individual for which expected utility is proportional to expected value of the lottery $\sum P_ix_i$. In this paper I make use of the utility function of Eq. (\ref{exp}). However the results of the papers remain valid also for other (but not all) forms of the utility function. In Section V I comment on this point while in the Appendix I consider the case of a logarithmic utility function, $u(x)=C\log(1+c x)$.

Consider the limit order placement as a problem of utility maximization. On one hand traders would prefer to place limit orders very far from the spread (i.e. with large values of $\Delta$) because this will increase trader's profit. On the other hand the larger the $\Delta$ the longer (on average) the time one has to wait until the limit order is fulfilled. So the trader has to find the right tradeoff between these two opposite choices. The right place where to place the limit order depends in general from three characteristics of the trader. First of all the optimal $\Delta$ depends on the patience of the trader. To model this aspect we introduce the time horizon $T$ of the trader, which is here defined as the time the trader is willing to maintain the limit order in the book before canceling it (if not matched). The second characteristic is the volatility $\sigma$ used by the trader to model price dynamics.  When the volatility is high, price fluctuates more and the trader places limit orders with larger values of $\Delta$. 
Finally the third characteristic is the utility function $u(x)$ of the trader. All else being equal,  more risk averse traders place limit orders closer to the best price.

\begin{figure}[ptb]
\rotatebox{-90}{\resizebox{2.2\columnwidth}{!}{%
\includegraphics{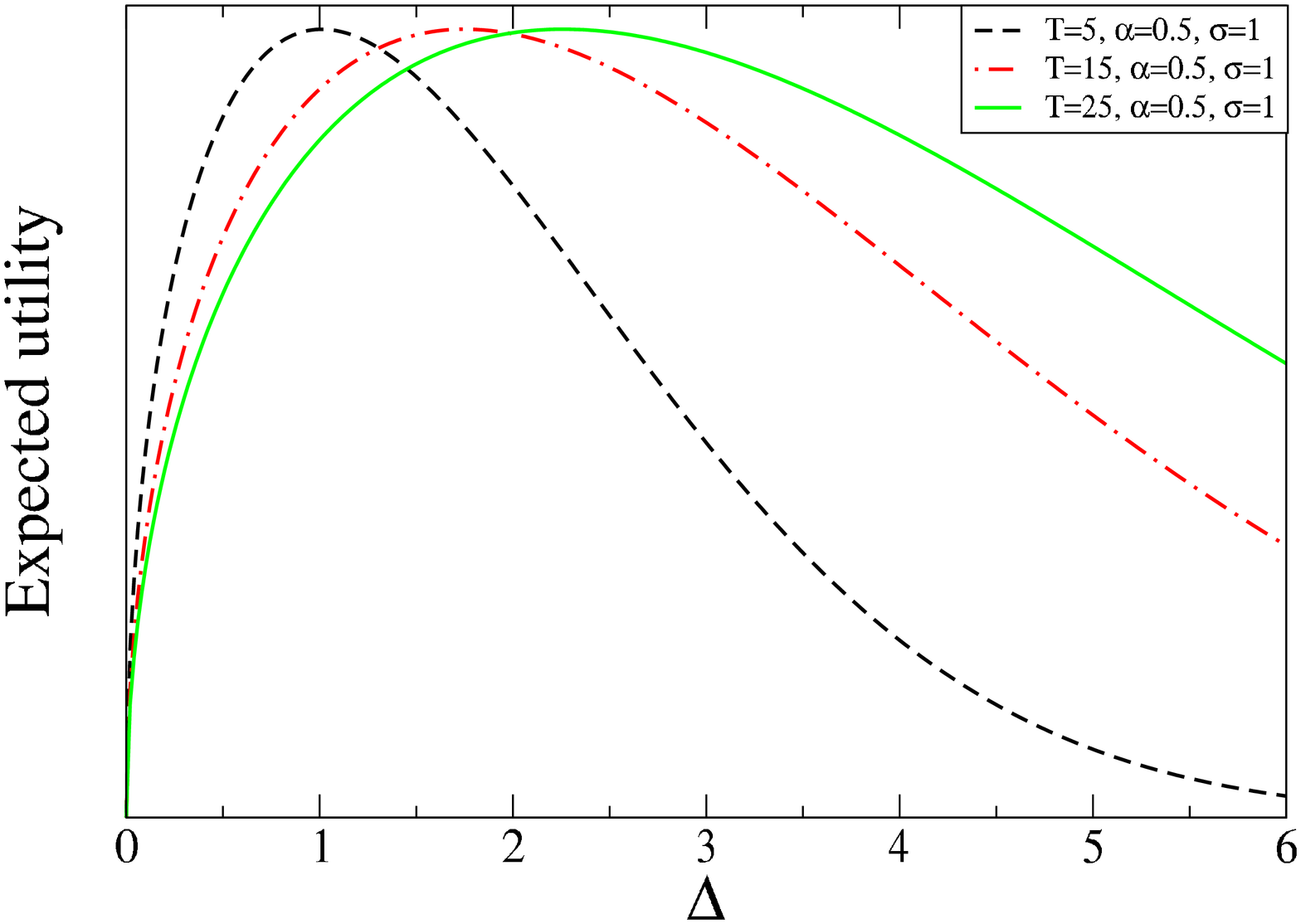}
\includegraphics{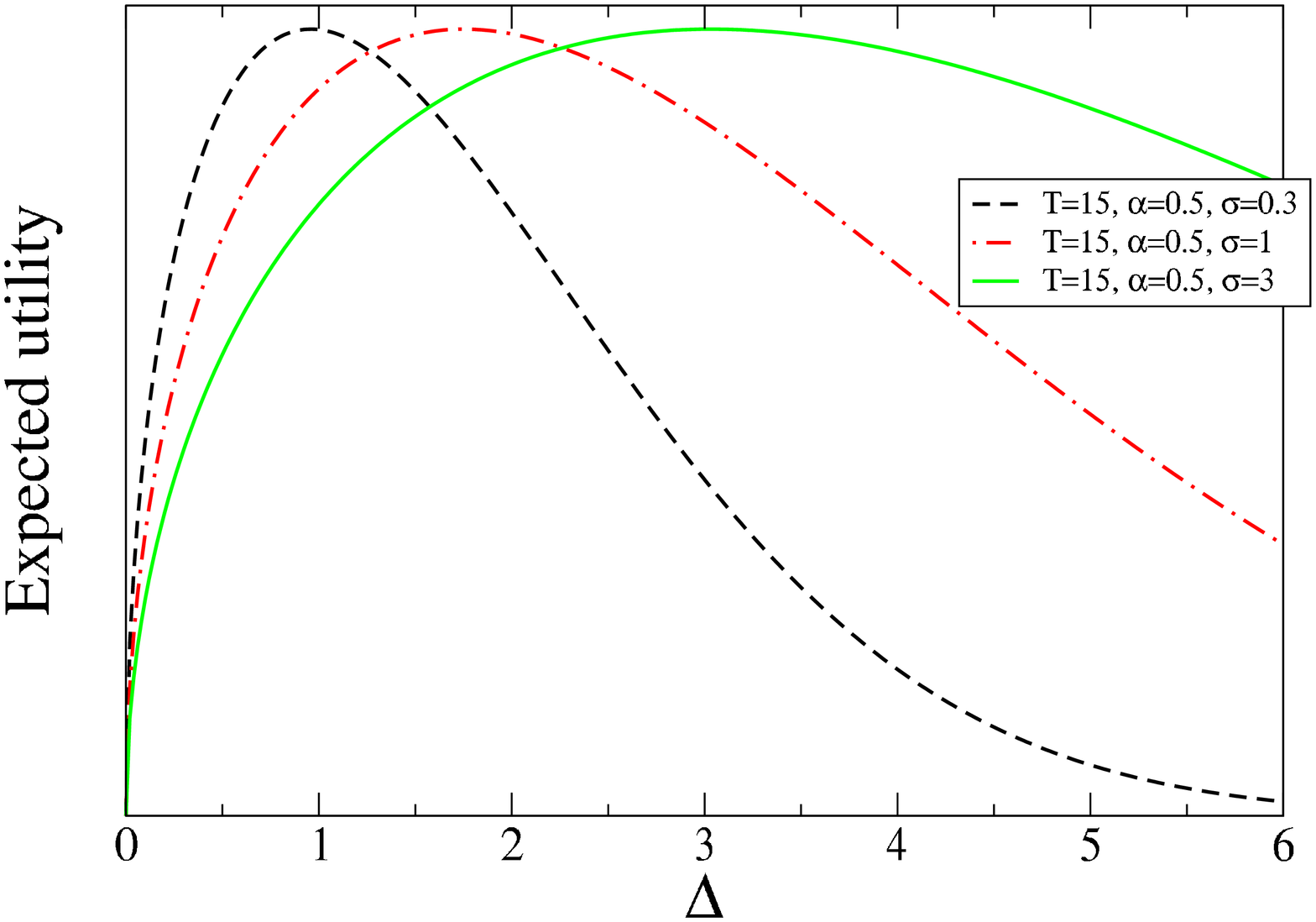}
\includegraphics{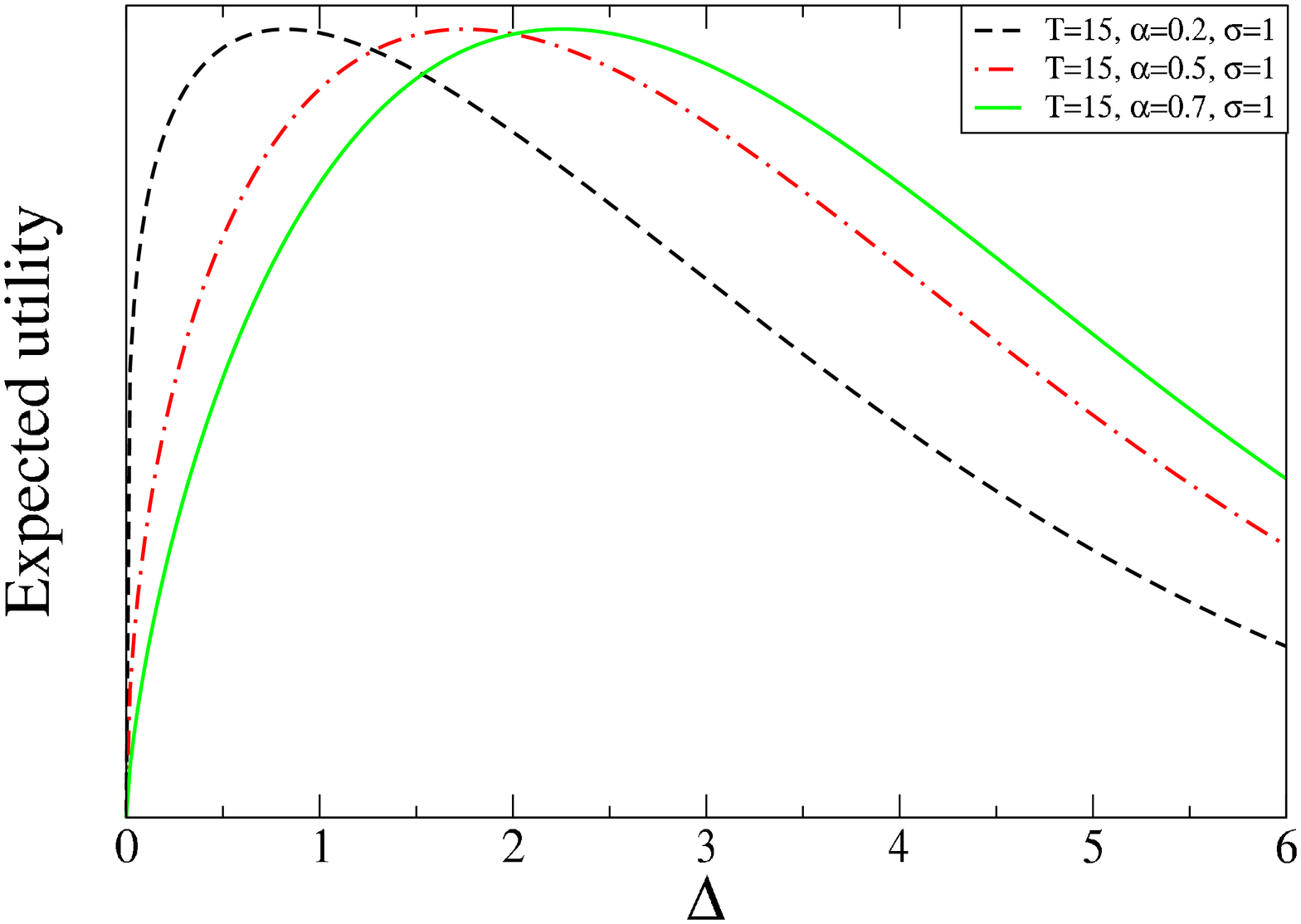}
}
}
\caption{Expected utility $U_{T,\sigma}(\Delta)$ of Eq.~(\ref{utility}) as a function of the distance $\Delta$ of the limit price from the corresponding best. The panels show the dependence of $U_{T,\sigma}(\Delta)$ from the time horizon $T$ (top), the volatility $\sigma$ (middle), and utility function parameter $\alpha$ (bottom). }
\label{utilityfig}
\end{figure}

The limit order placement problem can be formalized as follows. Given an time horizon $T$,  an utility function $u(x)$, a volatility $\sigma$, and a limit order price $\Delta$, the prospects facing the investor is $(\Delta,  P(\Delta,T,\sigma); 0, 1-P(\Delta,T,\sigma))$, i.e. she gains $\Delta$ with probability $P(\Delta,T,\sigma)$ if the limit order is matched in a trade in a time shorter than $T$ or she gains $0$ with  probability $1-P(\Delta,T,\sigma)$ if the limit order is canceled before the price reaches the limit order price. The trader's utility is
\begin{equation}
U_{T,\sigma}(\Delta)=P(\Delta,T,\sigma)~u(\Delta)+(1-P(\Delta,T,\sigma))~u(0)
\end{equation}
Without loss of generality we can set $u(0)=0$ because the expected utility can be rewritten as $P(\Delta,T,\sigma)(u(\Delta)-u(0))+u(0)$ and the value of $\Delta$ maximizing the utility $u$ is the same as the value maximizing $u-u(0)$.  
Thus to each limit order price it is associated a lottery with a different expected utility $U_{T,\sigma}(\Delta)=P(\Delta,T,\sigma)~u(\Delta)$ and the trader maximizes her utility by choosing the lottery with the largest expected utility. In other words trader maximizing utility  places limit orders at the value of $\Delta$ which maximizes $U_{T,\sigma}(\Delta)$, i.e. $\Delta^*\equiv argmax[U_{T,\sigma}(\Delta)]$.

In order to explicit the functional form of $U_{T,\sigma}(\Delta)$ we need to find the expression for  $P(\Delta,T,\sigma)$ which is the probability that the price random walk reaches the level $\Delta$ between $t=0$ when the limit order is placed and $t=T$ when the limit order is canceled. This probability is related with the first passage time distribution, which is the probability that a stochastic process reaches a given level $x$ for the first time at time $t$. Assuming that price performs a Brownian motion with diffusion rate $\sigma$, the first passage distribution is (see for example \cite{Oksendal})
\begin{equation}
f(\Delta,t)=\frac{ \Delta}{\sqrt{2\pi\sigma^2 t^3}}\exp[-\frac{\Delta^2}{2\sigma^2 t}]
\end{equation}
Thus the probability that the limit order is fulfilled in the interval $0\le t \le T$ is
\begin{equation}
P(\Delta,T,\sigma)=\int_0^T f(\Delta,t)~dt=erfc\left[\frac{\Delta}{\sqrt{2\sigma^2 T}}\right]
\end{equation}

Let us assume that trader has a constant relative risk aversion utility function with exponent $\alpha$ , then the function to be maximized is 
\begin{equation}
U_{T,\sigma}(\Delta) = erfc\left[\frac{\Delta}{\sqrt{2\sigma^2 T}}\right] ~C\Delta^\alpha
\label{utility}
\end{equation}
Figure~\ref{utilityfig} shows the dependence of $U_{T,\sigma}(\Delta) $ from the parameters.

The maximization of $U_{T,\sigma}(\Delta)$ is obtained by derivation with respect to $\Delta$. The maximum cannot be found analytically, but a simple scaling argument helps in understanding the solution. The equation $\partial_\Delta U=0$ can be rewritten as
\begin{equation}
C\Delta^{\alpha-1}\left[\alpha~ erfc[z]-\frac{2}{\sqrt{\pi}}ze^{-z^2}\right]=0
\end{equation} 
where I have set $z=\Delta/\sqrt{2\sigma^2 T}$. The solution of the equation obtained by putting to zero the term in square bracket can be obtained formally by solving the equation in $\alpha$
\begin{equation}
\alpha=\frac{2}{\sqrt{\pi}}\frac{z~e^{-z^2}}{erfc[z]}\equiv g(z)
\label{g}
\end{equation}
which gives $z=g^{-1}(\alpha)$ and with this position one obtains for $\Delta$ the optimal value
\begin{equation}
\Delta^*=\sqrt{2}g^{-1}(\alpha)\sigma T^{1/2}
\label{solution}
\end{equation}
This is the solution to the utility maximization problem and the rational trader should put the limit order at a distance $\Delta^*$ from the corresponding best price.  Eq.~(\ref{solution}) holds for power law utility functions. In the appendix I show that logarithmic utility function leads to the relation $\Delta^*\sim T^{1/2}/\log T$, i.e. there is a logarithmic correction to the square root relation. 

\section{On the origin of power law distribution of limit order prices}

In the framework of the utility maximization model described above, an homogeneous set of traders characterized by the same utility function (exponent $\alpha$), the same time horizon $T$ and the same volatility estimation $\sigma$ would place limit order always at the same distance $\Delta^*$ of Eq.~\ref{solution} from the best. This is at odds with what observed in real data, since, as we have discussed above, empirical investigations show that the limit order price distribution is very broad and probably compatible with a power law with a small exponent. We can then ask what is the origin of power law distribution of limit order prices according to our model. 

Since the optimal distance $\Delta^*$ of Eq.~(\ref{solution}) depends on $T$, $\alpha$, and $\sigma$, an heterogeneity in one of these three parameters would result in a distribution of $\Delta$ with a non-vanishing variance. I want to explore the possible role of the three parameters in being the most likely proximate cause of power law distribution of limit order prices. In real markets one expects an heterogeneity in all the three parameters, thus all contribute in principle to the distribution of limit price. I will consider the three cases in which two of the parameters are fixed and the third one is allowed to vary according to some distribution in order to see the separate effect of each parameter on the limit price distribution. In the Conclusions I argue that this way of proceeding is not a limitation.

{\it Heterogeneity in time horizon.} First of all let us consider the role of the heterogeneity of time horizon $T$. We assume that traders are characterized by a probability density function $P_T(T)$ of time horizon and that they are homogenous with respect to the utility function and to the volatility. From eq.~(\ref{solution})  the distribution of limit order price is then given by 
\begin{equation}
P_\Delta(\Delta)=P_T(T)\left|\frac{dT}{d\Delta}\right|\propto P_T(T)\Delta
\end{equation} 
The only way to have a power law distribution of limit order price like in Eq.~(\ref{pow}) is then to assume that the distribution of time horizon is asymptotically power law $P_T(T)\sim T^{-\zeta_T}$. In this case the limit order price are power law distributed with an exponent
\begin{equation}
\zeta_\Delta=2\zeta_T-1
\label{zetat}
\end{equation}
Under these assumptions the power law distribution of limit order prices is  the consequence of the power law distribution of time horizon. By using the empirical values for $\zeta_\Delta$ we can infer the value of $\zeta_T$. The value $\zeta_\Delta=2.5$ obtained by Zovko and Farmer \cite{Zovko} gives the value $\zeta_T=1.75$, whereas the value $\zeta_\Delta=1.6$ obtained by Potters and Bouchaud \cite{Potters} gives the value $\zeta_T=1.3$. It is very difficult to measure empirically the statistical properties of time horizon. However a recent paper by Borland and Bouchaud \cite{Borland} introduces a GARCH-like model obtained by introducing a distribution of traders' time horizons and the model reproduces empirical values of volatility correlation for $\zeta_T=1.15$, significantly close to our estimate. In an unpublished work by Vaglica {\it et al.} \cite{Vaglica} an estimate of the time horizon distribution is obtained for the Spanish Stock Exchange by computing the time a given institution is (statistically) maintaining its buy or sell position. The empirical distribution of time horizon is power law with an exponent $\zeta_T\simeq2.5$. In conclusion recent empirical results indicate the presence of a power law distribution of investment time horizons. The estimated value of tail exponent supports the view of the heterogeneity of time horizon as the proximate cause for fat tails in limit order prices.   

In order to check whether the result of Eq.~\ref{zetat} is valid only for power law utility function I consider also a logarithmic utility function $u(x)=C\log(1+cx)$. In the appendix I show that in this case 
\begin{equation}
P_\Delta(\Delta)\sim \frac{1}{\Delta^{2\zeta_T-1}~(\log \Delta)^{2\zeta_T-2}}
\end{equation} 
i.e., a part from a logarithmic correction, the limit order price distribution is still asymptotically power law with the same exponent as for power law utility function (Eq.~\ref{zetat}). 
The logarithm as well as its powers are slowly varying functions \cite{Embrechts}. In Extreme Value Theory the presence of slowly varying functions is unessential to the description to the asymptotic behavior of a function. More precisely all the functions $x^{\alpha}L(x)$, where $L(x)$ is any slowly varying function, belong to same Maximum Domain of Attraction.  As a consequence statistical estimators of the tail exponent which are based on Extreme Value Theory (for example the Hill estimator) give the same exponent $\alpha$ independently on $L(x)$. In other words the presence of the logarithmic correction is invisible to many tail exponent estimators. This means that utility functions different from Eq.~(\ref{exp}) may give the same scaling exponent $\zeta_\Delta$.

{\it Heterogeneity in utility.} The second hypothesis we want to test is whether power law distributed limit order prices can be explained by only assuming heterogeneity in utility function (i.e. in $\alpha$), assuming that $T$ and $\sigma$ are constant. Assuming a probability density function for $\alpha$, $P_\alpha(\alpha)$ and by using Eqs.~(\ref{g}) and (\ref{solution}), one can derive the formula for the distribution  of limit order price 
\begin{eqnarray}
P_\Delta(\Delta)=P_\alpha(\alpha)\frac{1}{\sqrt{2\sigma^2 T}}\\ 
\label{pdialfa}
\times \left[\frac{4 e^{-2z^2}z}{\pi erfc^2(z)}+\frac{2e^{-z^2}}{\sqrt{\pi}erfc(z)}-\frac{4e^{-z^2}z^2}{\sqrt{\pi}erfc(z)}\right]\nonumber
\end{eqnarray} 
where as before $z=\Delta/\sqrt{2\sigma^2 T}$. Since $g(z)$ of eq.~\ref{g} is a convex monotonically increasing function of $z$, large values of $\Delta$ are explained by large values of $\alpha$. The distribution of limit price is determined by $P_\alpha(\alpha)$ and in particular from its support. If for example the traders are risk averse, $\alpha<1$ and also the support of  $P_\Delta(\Delta)$ is bounded, meaning that there is a maximal value of $\Delta$ beyond which traders do not place limit orders. However it is possible to find a specific, yet quite artificial, distribution of the parameter $\alpha$ giving a power law distribution of limit order prices. Not surprisingly the utility distribution is power law, i.e. $P_\alpha(\alpha)\sim\alpha^{-\zeta_\alpha}$ in some interval of the exponent $\alpha$ governing the utility function of the traders. By expanding Eq.~(14) for large $z$ (i.e. large $\Delta$) I show that the asymptotic behavior of limit price distribution is $P_\Delta(\Delta)\sim \Delta^{-\zeta_\Delta}$ with $\zeta_\Delta=2\zeta_\alpha-1$. In order to match the empirical value for $\zeta_\Delta$ obtained by Zovko and Farmer one has to postulate a power law distribution of the parameter $\alpha$ with an exponent $\zeta_\alpha=1.75$, whereas the Potters and Bouchaud value gives $\zeta_\alpha=0.8$. Note that this last value is smaller than one implying that the support of $P_\alpha(\alpha)$ must be bounded from above. In conclusion the model is able to deduce the power law distribution of limit order price, but one needs to assume the presence of many investors with very large values of $\alpha$, i.e. unrealistically risk lover.

{\it Heterogeneity in volatility.} We now want to test the last hypothesis that changing volatility could be the proximate cause of power law in limit order price distribution. Volatility can be heterogeneous either because traders have different estimate of volatility or because volatility itself is fluctuating. We consider here this second possibility.
In this case, even if all the traders make the same estimate of volatility at a given time, the unconditional distribution of limit order price is broad because of the fluctuation of volatility. We have quoted above that Zovko and Farmer have observed a positive correlation between volatility and limit price. This correlation is captured by the solution of Eq.~(\ref{solution}). The point here is to check whether fluctuation in volatility could be able to determine a power law distribution of limit order prices. 
Since volatility $\sigma$ and optimal limit price $\Delta^*$ are proportional in Eq.~(\ref{solution}), in the framework of the proposed model the distribution of limit order price is the same as the distribution of volatility. This means that limit prices are power law distributed if volatility is power law distributed as $P_\sigma(\sigma)\sim \sigma^{-\zeta_\sigma}$. In this case $\zeta_\Delta=\zeta_\sigma$. In order to match empirical values for $\zeta_\Delta$  we should expect an exponent $\zeta_\sigma=2.5$ to explain Zovko and Farmer value and $\zeta_\sigma=1.5$ to explain Potters and Bouchaud value. Many recent measurement of volatility distribution find a power law tail, but the fitted exponent is too large when compared to these values. For example Liu {\it et al.} \cite{Liu} found a tail exponent $\zeta_\sigma$ slightly dependent on the time interval used to compute volatility and ranging between $4.06$ to $4.38$. Miccich\`e {\it et al.} \cite{Micciche} fitted a different proxy of volatility finding an exponent $6.27$. Finally the fit reported in the book by Bouchaud and Potters \cite{Bouchaudbook} gives an estimate $\zeta_\sigma=7.43$. Although the proxy used to estimate volatility can influence significantly the fitted value of $\zeta_\sigma$, it is quite clear that the empirical values are not consistent with the value needed to explain limit order price distribution. 

In conclusion, in the framework of the present model heterogeneity in  time horizon $T$ appears to be the most likely explanation of power law distribution of limit order prices.

\section{Empirical analysis}
It is quite difficult to assess empirically which of the variables $T$, $\sigma$ and $\alpha$ (or the utility function) is determinant in explaining the fat tails of limit order price distribution. This is due to the fact that $T$ and $\alpha$ are unobservable variables and volatility $\sigma$ can be measured in many different ways. Hence the purpose of the present empirical analysis is to convince qualitatively that the explanation given above of the power law distribution of limit order prices is plausible. I investigate this problem by considering the data of the London Stock Exchange (LSE) in the period May 2000 - December 2002. 

\begin{figure}[ptb]
\rotatebox{-90}{\resizebox{1.6\columnwidth}{!}{%
  \includegraphics{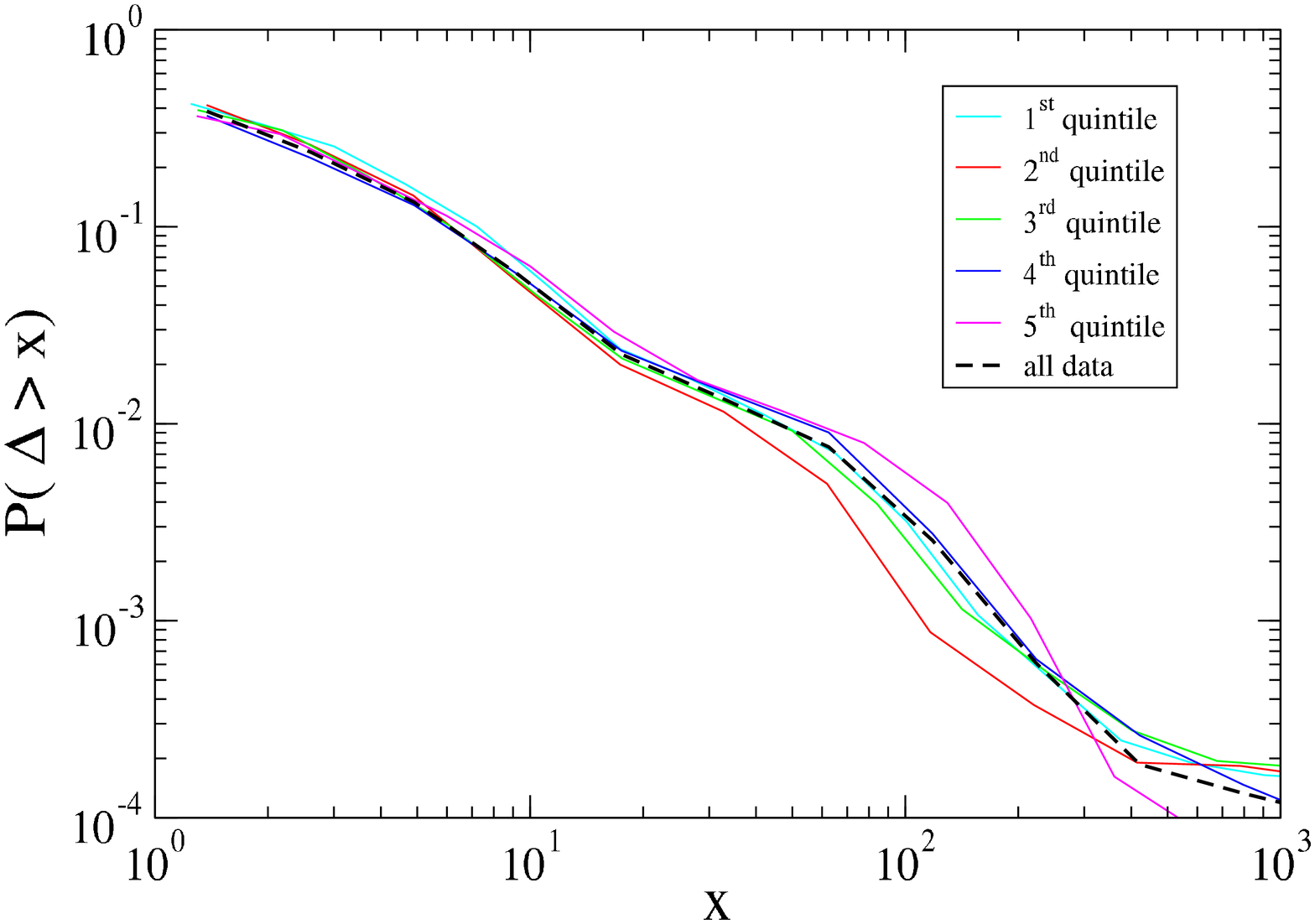}
\includegraphics{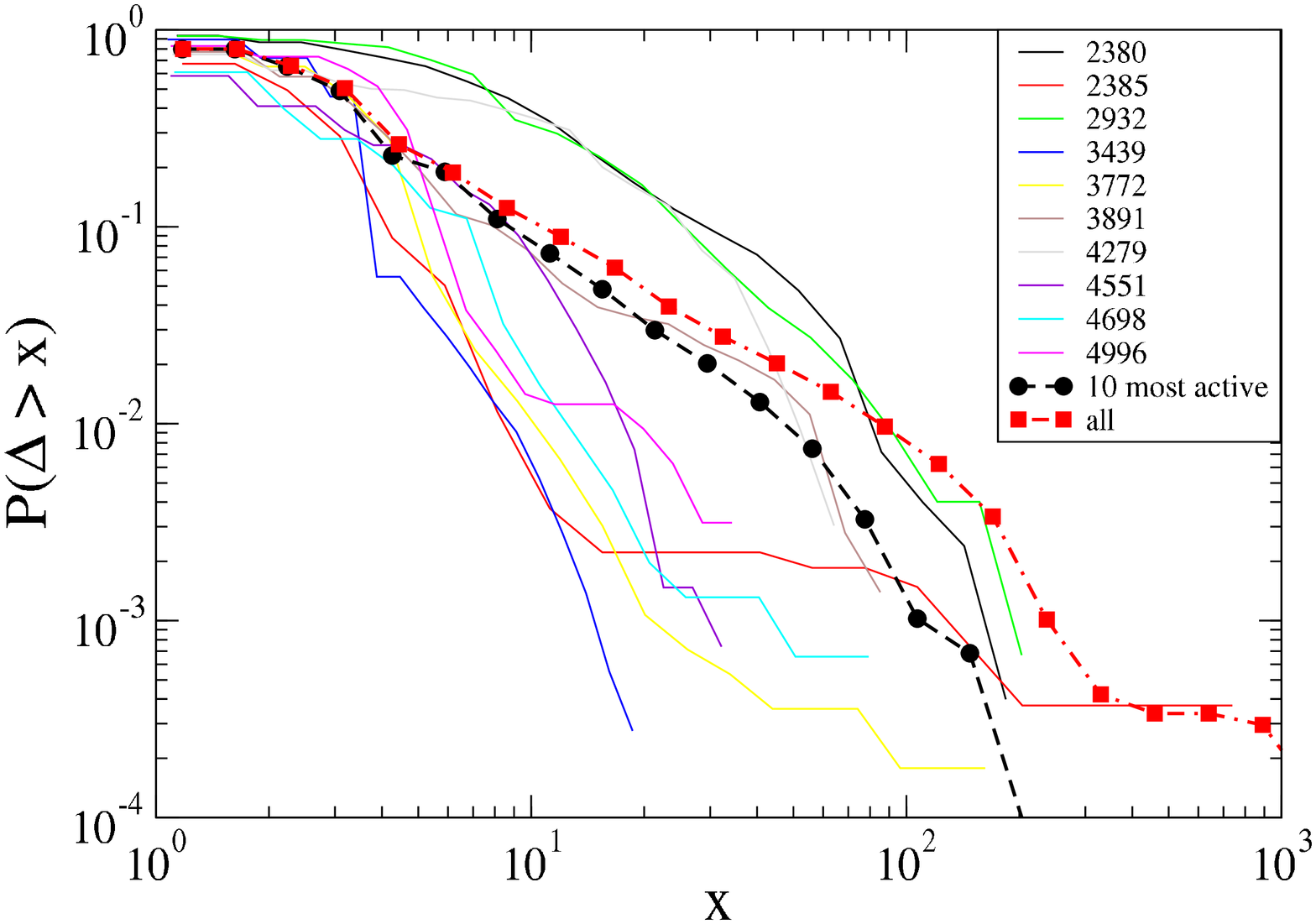}
}
}
\caption{ Top. Cumulative distribution of sell limit order price (in pence) for Astrazeneca traded at LSE in the period May 2000- December 2002. The black dashed line is the unconditional distribution whereas the solid lines are the distribution of limit order price conditional to the volatility in the day when the order was placed. The trading days are divided in five quintiles according to the volatility value. Bottom. Cumulative distribution of sell limit order price (in pence) for the $10$ most active institutions placing limit orders in the stock Astrazeneca during October 2002. The number in the legend is the institution code in the LSE database. Blue circles describe the cumulative distribution for the pool of the $10$ most active institutions, whereas red squares describe the cumulative distribution for the pool of all the $104$ institutions.}
\label{volatilityfig}
\end{figure}

In order to assess the role of volatility fluctuations, in top panel of Figure~\ref{volatilityfig} I compare the unconditional distribution of $\Delta$ with the limit order price distribution conditional to the volatility in the day when the limit order was placed. I divide the $675$ days of the sample in five quintiles according to the volatility value and then I plot the limit price distribution for the different quintiles. It is seen that $P(\Delta|\sigma)$ is weakly dependent on $\sigma$ strengthening the conclusion that  volatility fluctuations are unable to explain power law distribution of limit order prices. This conclusion is also supported by a direct measurement of $\zeta_\Delta$ by using the Hill's estimator \cite{Embrechts}. The value obtained for the five subsets ranked in increasing volatility are $\zeta_\Delta=1.46\pm 0.07$, $1.50\pm 0.07$, $1.51\pm 0.07$, $1.59\pm0.07$, and $1.42\pm0.05$. Finally it is worth noting that I obtain the same weak dependence of $P(\Delta)$ when the conditioning is made on the mean volatility in the five previous days rather than on the volatility in the same day when the limit order was placed. 

As said above, testing for a dependence of limit order placement from the agents is very difficult due to lack of data. However the LSE database allows us to track the
actions of individual institutions through a numerical code which
identifies the institution. For privacy reasons the code is different
for different stocks and it is reshuffled each calendar
month. Therefore our analysis will be limited to a single trading
month.  I consider as a case study the stock Astrazeneca in October 2002 ($1.1~10^5$ limit orders). There are $104$ active institutions for the considered month but the activity distribution is quite skewed. In fact the $10$ most active institutions are responsible for more than $80\%$ of limit orders. 
Bottom panel of Figure~\ref{volatilityfig} shows the limit order price distribution for the $10$ most active institutions. The figure shows that there is a large variation of the form of the distribution suggesting that the large heterogeneity in institution's limit order strategy could be the driving factor of power law limit order price distribution. It is worth pointing out that this empirical analysis cannot distinguish whether the heterogeneity among different institutions is due to an heterogeneity in time horizon $T$ or in the utility function. However the argument above indicates that heterogeneity in utility function should play a minor role. In conclusion the comparison of the two panels of Fig.~\ref{volatilityfig} indicates that heterogeneity in traders' behavior is much more important than fluctuation in volatility in explaining limit order price distribution.

\section{Conclusions}

I have shown that treating the limit order placement as an utility maximization problem gives insight on the origin of the power law distribution of limit order prices. The main conclusion is that the heterogeneity in time horizon is the proximate cause of this power law. Empirical analysis suggests that this could be the correct explanation.
One could naively expect that all the three parameters $T$, $\alpha$ and $\sigma$ can simultaneously contribute to the fat tailed distribution of limit order price. However power law distributions satisfy nice aggregation properties (see, for example, \cite{Gabaix}). If a variable $y$ is the product of $n$ independent asymptotically power law distributed variables with different tail exponents, $y=\prod_i x_i$, then $y$ is asymptotically power law distributed with a tail exponent equal to the minimum tail exponent of the variables $x_i$. This argument shows that even if all the variables contribute to the distribution of $\Delta$, only one determines its tail exponent and our discussion above indicates that the most  likely candidate is the time horizon $T$.

The conclusion I draw on the origin of the fat tail of limit order price distribution is based on the choice of the utility function. In this paper I consider power law and logarithmic utility function showing that in both cases heterogeneity in time horizon seems to be the key variable.
This conclusion may not be true for other utility functions. For example the exponential utility function $u(x)\propto (1-\exp(-a x))$ seems to behave in a different way. The optimal limit order price cannot be found analytically, but numerical analysis suggests that in this case $\Delta^*\propto \log(T)$ rather than $\sqrt{T}$. As a consequence the time horizon argument leads to $P_\Delta(\Delta)\propto \exp(-(\zeta_T-1)\Delta)$ rather than a power law distribution. The reason for this behavior is the extreme risk aversion of an investor with an exponential utility function which forces the investors to place limit orders very close to the best price, even when the investor has a very long time horizon.  

Finally, in this paper we have considered limit orders placed inside the book, i.e. sell (buy) limit orders with a price higher (lower) than the ask (bid). In other words we considered only positive $\Delta$. Recent empirical analysis \cite{Mike,Ponzi} shows that also prices of limit orders placed inside the spread ($\Delta<0$) are power law distributed with an exponent close to the one for $\Delta>0$. One is tempted to extend the utility maximization argument to describe limit orders inside the spread. However limit orders in the spread are usually placed by agents with a very different strategy with respect to traders placing orders for large positive $\Delta$. Other risk factors, such as adverse selection, enter in the decision process. Even if the utility maximization approach can be useful in tackling this problem, the case $\Delta<0$ is outside of the scope of the present paper and will be considered elsewhere.

I acknowledge support from the research project
MIUR 449/97 ``High frequency dynamics in financial markets" and
from the European Union STREP project
n. 012911 ``Human behavior through dynamics of complex social
networks: an interdisciplinary approach.''.
I wish to thank Doyne Farmer for providing the LSE data with the identity code for the institutions and for useful discussions. I also thanks Rosario Mantegna and Michele Tumminello for useful discussions.

\section{Appendix: Logarithmic utility function}

We consider here the case of a logarithmic utility function $u(x)=C\log(1+c x)$. The expected utility is 
\begin{equation}
U_{T,\sigma}(\Delta) = erfc\left[\frac{\Delta}{\sqrt{2\sigma^2 T}}\right] ~C\log(1+c\Delta)
\end{equation}
By setting to zero the derivative $U_{T,\sigma}(\Delta)$ with respect to $\Delta$ and considering large $\Delta$, one obtains
\begin{equation}
h(z)\simeq c\sqrt{2\sigma^2 T}
\end{equation}
where $z=\Delta/\sqrt{2\sigma^2 T}$ and
\begin{equation}
h(z)\equiv\frac{\exp\left[\frac{\sqrt{\pi}}{2}\frac{e^{z^2}~erfc(z)}{z}\right]}{z}
\end{equation}
Thus the optimal limit order price is
\begin{equation}
\Delta^*=\sqrt{2}\sigma T^{1/2} h^{-1}(c\sqrt{2\sigma^2 T})
\label{solutionlog}
\end{equation} 
As for the power utility function the optimal limit price $\Delta^*$ is expressed implicitly in terms of an inverse function. The main difference with Eq.~(\ref{solution}) is that the argument of $h^{-1}$ now contains the variables $T$ and $\sigma$ and it is not immediately obvious what is the asymptotic behavior of $\Delta^*$ for large $T$. In order to answer this last question we need to study the asymptotic behavior of $h^{-1}$. The function $h(z)$ diverges in $z=0$ and is monotonically decreasing. Thus the asymptotic behavior of $h^{-1}$ is determined by the behavior of $h(z)$ around $z=0$ which is
\begin{equation}
h(z)\simeq\frac{1}{e}\frac{\exp\left[\frac{\sqrt{\pi}}{2z}\right]}{z}
\end{equation}
We prove that $h^{-1}(x)\sim1/\log(x)$. In fact
\begin{eqnarray}
\lim_{x\to \infty}\log(x)h^{-1}(x)=\lim_{z\to 0}\log (h(z)) z\\\simeq \lim_{z\to 0} \log(\frac{1}{e}\frac{\exp\left[\frac{\sqrt{\pi}}{2z}\right]}{z})z=\frac{\sqrt{\pi}}{2}
\end{eqnarray}
This result can be used to determine the asymptotic behavior of $\Delta^*$. From Eq.~\ref{solutionlog} we get
\begin{equation}
\Delta^*\sim \frac{2\sqrt{2}}{\sqrt{\pi}}\frac{\sigma T^{1/2}}{\log(c\sqrt{2}\sigma T^{1/2})}\sim \frac{T^{1/2}}{\log{T}}
\end{equation}
In conclusion the optimal limit price for a logarithmic utility function scales with $T$ in the same way as for the power utility function, except for a logarithmic correction. 

Finally we consider what is the distribution of limit order price $\Delta$ under the assumption of logarithmic utility and of power law distribution of time horizon, $P_T(T)\sim T^{-\zeta_T}$. We know that $P_\Delta(\Delta)=P_T(T)\frac{dT}{d\Delta}$. Since $\Delta\sim \sqrt{T}/\log T$, it is $dT/d\Delta=(d\Delta/dT)^{-1}\sim \sqrt{T}\log T$. Thus implicitly 
\begin{equation}
P_\Delta(\Delta)\sim \frac{\sqrt{T}\log T}{T^{\zeta_T}}
\end{equation}
It can be shown that 
\begin{equation}
P_\Delta(\Delta)\sim \frac{1}{\Delta^{2\zeta_T-1}~(\log \Delta)^{2\zeta_T-2}}
\label{finale}
\end{equation}
In order to prove this result we consider the limit of 

\noindent $\Delta^{2\zeta_T-1}~(\log \Delta)^{2\zeta_T-2} P_\Delta(\Delta)$ for $\Delta\to\infty$ and we perform the limit by transforming it in a limit $T\to\infty$. We show that this limit is a finite non-vanishing constant, proving Eq. (\ref{finale}).

\end{document}